# Jo Wilder and the Capitol Case: A taxonomy of uses for a historical inquiry game in 4th grade Classrooms in Wisconsin

Peter Wardrip, David Gagnon, James Mathews, Jen Scianna

Abstract: In this paper, we study the various ways 3rd-5th grade educators in Wisconsin utilized Jo Wilder and the Capitol Case, a historical inquiry game, as part of their classroom instruction. The 15 educators involved in the study were all grade school teachers in Wisconsin who took part in the "Doing History Fellowship" program, a professional development opportunity offered by the authors, designed to increase their understanding of historical inquiry instruction and game-based learning. As part of the program, the educators planned and implemented the game within their own classroom context and reported their results back to the authors and other educators. Through their reports, surveys and semi-structured interviews we discovered the educators were motivated by five distinct instructional purposes, which influenced how the game was integrated into their curriculum. In this paper, we name and describe these five purposes. We see these findings as useful insights into how educators think about games and how educational video games and corresponding professional development activities may be designed in the future.

Keywords: Game-based learning, Instruction, K-12 Education, Social Science Instruction

**Introduction**

Advocating for the use of video games in educational contexts is nothing new. In 2006, the Federation of American Scientists (FAS) issued a widely publicized report stating that games as a medium offer powerful affordances for education. More recently, a meta-analysis of research demonstrated that digital games significantly enhanced student learning compared with non-game environments (Clark et al., 2016). Yet, the same review noted there is a great deal of variation in both enactment and outcome. While recent studies have mainly focused on the affordances of games for learning in educational contexts (e.g De Freitas, 2018), there are emerging studies focusing on pedagogical approaches that support the use of games in classrooms (Nousiainen, Kangas, Rikala, Nousiainen, 2018; Shah & Foster, 2015). These studies, like those exploring teacher enactment of innovations and policies (e.g. Cohen, 1990), recognize the important role teachers play in implementing new practices. This study we present here explores the implementation of a particular educational digital game and seeks to identify the forms of teacher enactment when integrating a historical inquiry game called *Jo Wilder and the Capitol Case* into a larger instructional unit.

As games grow in their use as teaching tools, it is important to acknowledge that like other curricular elements, they are enacted in classrooms in many different ways. Some have referred to these differing enactments as the "implementation paths" teachers take as they utilize educational technologies across a trajectory or set of phases, whereby they progress from initial to more effective use of a particular learning tool (Bielaczyc & Collins, 2006). Others have highlighted the ways in which teachers make curricula their own as they modify and adapt different materials and other resources to meet specific learning goals and address the needs and barriers they encounter (Shah & Foster, 2015; Squire, et al., 2003). By furthering our understanding of how teachers enact games, we can help those of us in the field provide enhanced implementation support materials, offer more responsive professional development opportunities, and create better informed policies that support the broader use of digital games in the classroom.

**Literature Review**
*Game-based Learning in Classrooms*
Digital video games have been used and formally studied over the last several decades across a wide range of educational contexts and topics. While there are critiques of digital games (e.g., Mayer, 2014), the overwhelming evidence is in favor of their usefulness. Educational games have been empirically shown to enhance students' understanding of targeted concepts, as well as skills (e.g., Clark, Tanner-Smith, & Killingsworth, 2015; D'Angelo et al., 2014), and have also been linked to increased motivation and interest in learning (National Research Council 2011; Metcalf et al., 2014). When well designed, games allow players to affect, experiment and observe consequences in the simulated world in ways that are representative of the real world. Thus, games can produce situated learning experiences tied to disciplinary-specific concepts and language (Gee, 2006a; Squire, 2006). James Gee (2006b) posits that games can situate players perceptually, narratively, and socially in a way that leads to new ways of thinking, knowing, and being in the world (Shaffer, Squire, Halverson, & Gee, 2005). Because players' actions within games are mediated, reducing danger and cost while expanding or compressing time, perspective, etc., games provide a useful means of teaching complex concepts, particularly ones that cannot be experienced directly (Kamarainen, Metcalf, Grotzer, & Dede, 2015). Games allow learners to take on virtual roles and become embedded within a story as an active agent experiencing intentionality (they choose where the story should go), legitimacy (have authentic ways of influencing the virtual world), and consequentiality (effects follow their actions) (Barab, Gresalfi, & Ingram-Goble, 2009). In some instances, games can be developed as a way for learners to encounter and adopt an entire epistemic frame (Shaffer, 2006a; Shaffer, 2006b) that envelops the values, skills, identities, and discourses of a professional community.

More and more, schools and educational programs are taking up game-based learning. As part of their 2013 national survey, the Joan Ganz Cooney Center surveyed 700 teachers about how they use digital games in the classroom. They found that nearly three-quarters (74%) of K-8 teachers report using digital games for instruction. Four out of five of these teachers say their students play at least monthly, and 55% say they do so at least weekly. Digital game-using teachers also say they're using games to deliver content mandated by local (43%) and state/national curriculum standards (41%), and to assess students on supplemental (33%) and core knowledge (29%). The 2015 Speak Up survey from Project Tomorrow that surveyed K-12 educators, reports that only 47% of K-12 teachers use games, but highlight a significant increase from the 2010 result of 23%.

*Enactment of Game-based Learning*
Enactment models of instruction are often used to investigate the degree to which classroom events are congruent with the often overly ambitious instructional practices embedded in particular educational tools (Schneider, Krajcik, & Blumenfeld, 2005). While recognizing that teachers are professionals who need support with carrying out instruction, these models often explore the difference between the planned, intended and enacted versions of curriculum (Stein, Remillard & Smith, 2007). The planned use of curriculum is usually stated by the curriculum designer and often includes prescribed steps or instructional purposes. The intended use refers to how a teacher envisions implementing a curriculum or tool, including modifications and changes that may differ from the designers intention. Finally, how curriculum is actually enacted can differ from intention based on a variety of contingent classroom factors, including local constraints and pedagogical goals (Squire, et al., 2003; Stein et al., 2007).

How teachers implement educational innovations, especially when we understand that their enactment is contingent on a whole host of factors, is a consequential consideration for educational designers and professional development providers. These factors not only include the affordances of the innovation, but also the affordances of the classroom environment, teachers' current instructional practice and their beliefs and attitudes about their students and learning (Clarke & Hollingsworth, 2002). If we acknowledge the role teachers play in adapting curriculum materials to meet their local needs and align with their goals, then understanding the ways in which implementation happens can be useful (Squire et al., 2003; Squire, 2005; Barab & Luehmann, 2004). In the context of game-based learning, the enactment of a particular curriculum can be the negotiated result of a variety of ways in which a particular game can be utilized within a classroom. This acknowledgement of variation in implementation positions teachers as customizers who modify and remix curricular materials in order to meet the needs of their classroom, rather than simply recipients of pre-designed curricula. It also represents a larger tension that exists in educational technology between scale and localization (McMillan-Culp & Honey, 2000) and highlights the need for curricular tools and materials to have adaptive flexibility (Fishman & Kraijcik, 2003; Scharwtz, Lin, Brophy, Bransford, 1999; Squire, et al, 2003).

As an entry point for understanding teacher enactment of games, we specifically look at how teachers perceived the instructional purpose of a specific educational game at various points of their planning and implementation and explore how this shaped how it was integrated into the curriculum. Research has documented the role of teachers critiquing instructional materials, including materials' purposes or goals, as being a first step toward adapting the materials for local instructional use (Barab & Luehmann, 2003; Davis, 2006). This is a process of developing localized purposes, which are the starting point for actual enactment.

**Aim of Study**
In this paper we examine how a group of teachers used *Jo Wilder and the Capitol Case*, a point-and-click adventure game that requires players to engage in historical inquiry practices to locate, analyze and identify a series of artifacts. In doing so, we address the following questions:

- RQ1: How did teachers enact *Jo Wilder* in their classroom?
- RQ2: How did teachers' enactments of *Jo Wilder* integrate with a broader project?

Through data collected from a group of teachers involved in the game implementation fellowship, we build an emerging framework that describes the various instructional purposes that shaped how the game was used. In addition to further expanding our understanding of how games are implemented in classrooms, we see this framework as useful lens for developing professional learning experiences for teachers interested in using games in their teaching. We also see it as a useful framework for educational game designers as they develop additional resources to support teachers' use of their games and consider strategies for getting teachers to envision how particular games might be used to support teaching and learning.

**Methodology**
*Background: Designing Jo Wilder and the Capitol Case*
*Jo Wilder and the Capitol Case* is an online point and click adventure-style game that requires players to engage in historical inquiry practices to locate, analyze and identify a series of artifacts before they are displayed as part of a public exhibition. The game is designed for use in 3rd-5th grade and is aligned with Social Studies and Disciplinary Literacy Standards found in the Wisconsin Academic Standards.

Point and click adventure games are narrative-based games where the player commonly takes on the role of a protagonist who explores the world while completing a series of quests or solving a collection of puzzles. The player uses a computer mouse (or taps on the screen) in order to move around the game world and interact with people and objects that appear on the screen. The quests and puzzles are typically coupled with the narrative and players are required to complete them in order to advance the storyline. Some examples of this genre include *The Secret of Monkey Island*, *Phoenix Wright: Ace Attorney*, and *Syberia*.

In *Jo Wilder*, players take on the role of Jo, a middle school-aged girl whose grandpa works at the state history museum. The main antagonist in the game, Wells, is an aspiring curator who is trying to undermine Jo's grandpa and push him into retirement. The game opens just as a new exhibit about Wisconsin's history is being planned for the museum. Because of his desire to be the lead curator for the exhibit, Wells says and does things aimed at discrediting and sidelining Jo's grandpa. In his eagerness to demonstrate his productivity and shine in front of the museum's boss, however, Wells takes shortcuts and sloppily misidentifies several artifacts. It is up to Jo to highlight Wells' incompetence, thereby saving the museum from embarrassment and securing her grandpa's job. In order to do this, she needs to put her own historical inquiry skills into use; locating and interpreting artifacts, corroborating clues and producing sound, evidence-based arguments.

From a game design perspective, *Jo Wilder* leverages the affordances of the point and click adventure genre to produce a narrative hook and story structure that motivates players to engage in the disciplinary practices associated with *doing* history. While the historical content and themes in the game (e.g., women's suffrage and environmental activism) are aligned with academic standards, the main focus is on providing an experience whereby students are required to use historical practices such as close-reading, analyzing historical artifacts, corroborating details, and making evidence-based arguments in order to progress through the game. Additionally, while the main text of the game is delivered using vernacular language (and Tier 1 and Tier 2 vocabulary), specialist or domain-specific language (and Tier 3 vocabulary) is also introduced in a situated manner (Beck, McKeown & Kucan, 2002; Gee, 2004), as are occupations (e.g., archivist), tools (e.g., microfiche reader) and places associated with the history profession (e.g., archives).

*Jo Wilder* was developed through a partnership between UW-Madison's Field Day Lab and Wisconsin Public Television (WPT). In addition to our internal design teams, we recruited several other collaborators to provide guidance and insight related to different components of the game and its subsequent use as a teaching tool. This included content experts (i.e., professional historians, archivists and curators), pedagogical content experts (i.e., teachers and other professionals knowledgeable about current research and practices associated with teaching history), and elementary school teachers. To guide the development of the game our team implemented two Teacher Fellowship programs that involve teachers in the design process from early ideation to playtesting. The first Teacher Fellowship focused on completing a needs assessment and developing initial ideas for the game, while the second one engaged teachers in co-designing and testing various versions of the game with their students.

Between October 1, 2018 (when the game was launched) and September 30, 2019 (the time of writing), *Jo Wilder and the Capitol Case* was played 76,310 times nationally. The vast number of the game's users access the game online with no contact with Field Day or WPT staff. 81% of the users access the game via BrainPOP.com, 18% access the game though wpteducation.org and the remaining 1% of users find the game from a link on a class website or social media. On BrainPOP.com, the game is accompanied by a list of related national standards, an online lesson plan, sample reflection questions and guidelines for setting cooperative gaming expectations. From wpteducation.org the game is accompanied by instructions for game play, a downloadable standards document and a downloadable document for guiding reflection questions. No usage analytics are available for these supplemental resources.

*Doing History With Games*: Implementation Fellowship
After producing and publicly releasing *Jo Wilder* we turned our attention toward understanding how teachers implemented it in a variety of classroom contexts. To achieve this goal, we developed the *Doing History with Games Fellowship.* The fellowship, which took place over five months, was designed to provide teachers with background about *Jo Wilder* and support them in integrating the game within their classroom to teach historical thinking. The Fellowship, which was influenced by key recommendations from the PD literature, (e.g. Penuel et al, 2007). It included an opening workshop where we introduced the game and framed it within the larger context of history education. As part of the workshop participants met with archivists and educational outreach staff from the Wisconsin Historical Society to learn about the professional practices of historians, visited the state archives to learn about archival research, participated in several historical object study activities, and began planning how they would use the game in the coming months. After the workshop teachers were asked to finalize their implementation plan and enact it in their classroom. They were also asked to complete a report where they reflected on the implementation, including how they would use it differently in the future. Finally, we had a closing workshop where teachers shared the results of their implementation and discussed ideas for how to promote the game with other teachers.

Following the workshop, the teachers iterated on their project plans, optionally met online with Field Day staff to discuss their plan, implemented the game and documented their implementation in preparation for the second workshop. During the second workshop, they again met with Wisconsin Historical Society staff, heard about a historical community-based education project and reported on the implementation of the game. While we introduced teachers to models and activities related to historical inquiry we intentionally encouraged them to develop implementation plans that aligned with their local needs.

Participants and Settings

The 15 teachers who participated in the Fellowship were all fourth-grade teachers who taught social studies as part of their curriculum, but also taught other content (like English Language Arts, Science and Math). The teachers applied to be part of the Fellowship and were selected based on an expressed interest in integrating games to support learning in their classes. The teachers represented a range of rural, suburban and urban schools and varied in terms of their experience integrating games and historical inquiry into their teaching.

Data Collected

In this study, we drew from several sources of data. A list of the data as well as the general timeline for the data can be seen in table 1. First, we collected implementation plans from teachers. The implementation plans were online questionnaires that included open ended questions about the intentions the teachers had for the enactment of *Jo Wilder* within an instructional unit in their classroom. Second, we collected implementation reports from the teachers that consisted of open-ended questions about how the implementation actually unfolded with their class.

Third, participating teachers created powerpoint presentations to describe their implementation with the other fellowship participants. These powerpoint presentations were presented in person and included images and text to describe their implementation. Finally, a subset of five participants were interviewed using a semi-structured protocol (Rubin & Rubin, 2005). The interviews were administered before the implementation reports were completed. The interviews served to provide specific examples of how the implementation was taking place and allowed our team to ask clarifying questions related to shifts in the teachers thinking as they moved from planning to implementation.

Table 1: Teacher Fellowship Program Sequence

| **Data Source** | **Timeline** |
| --- | --- |
| Implementation Plan | October-November |
| Semi-structured Interviews | December-January |
| Implementation Report | February |
| Project Presentations | February |

Data Analysis

The analytic process began during data collection. Throughout the fellowship, three researchers engaged in weekly discussions about the data to test conjectures, clarify terms and begin the analytical framing of the implementations. These discussions also enabled us to clarify the coding categories and continuously seek counter-examples in the building of our framework (Corbin & Strauss, 2008). As the data were collected, we were able to add additional questions or collect additional data during subsequent waves of data collection. For example, we were able to ask questions in the interviews based on what was read in the implementation plan for particular teachers, such as having the teachers explain different changes they made. Our analyses in developing this initial framework were inductively developed from the data and informed by the perspective that the teachers may enact the use of games in a variety of ways. Therefore, this analysis was not "goal free" (Scriven, 1991), but rather guided by an assumption that the data from the teachers would suggest ways in which they used the game to address a particular need or goal in their classroom.

      We sought credibility in our analysis through a number of strategies (Lincoln & Guba, 1986). First, we sought to maintain methodological consistency through our data collection and analysis (Morse, Barrett, Mayan, Olson, & Spiers, 2002). Therefore, our data and analysis were aligned with our research question and theoretical perspective on teacher enactment. This was not intended to constrain our analytic process but to ensure a "trustworthiness" (Lincoln, 1995) in that our point of inquiry, analytic approach, and analysis were carried out systematically and as intended. Finally, the researchers critically discussed the analyses on an ongoing basis. This not only supported a critical eye to the analytic process, but also to the claims that came from the analysis.

**Results**

Our analysis of the data suggests the teachers enacted the game to meet five different instructional purposes. In table 2, we present each of these instructional purposes along with a brief description. Note that while some teachers focused on only one of the instructional purposes, others combined the purposes as part of their curriculum enactment.

Table 2: Instructional Purposes of Games in the Classroom

| **Instructional Purpose** | **Description** |
| --- | --- |
| Game as Reading Text | The game was used as part of reading instruction, for the purpose of improving reading skills. |
| Game as Content Container | The game was used for its factual historical content such as names and dates. |
| Game as Practice Space | The game was used as a context for practicing historical inquiry skills. |
| Game as Model | The game was used as a simplified example to help students think about how and when historical practices are applied. |
| Game as a Launchpad | The game was used to prepare students for an out-of-game historical inquiry project. |

**Games as reading text**

One instructional purpose we saw during this study was the use of the game as a text to promote reading literacy. This is not surprising given that the game includes 1,000 lines of written text. As part of this instructional purpose, we saw the game used both as a way to help develop students' decoding and comprehension skills and as a context for reinforcing and practicing reading strategies used elsewhere as part of previous classroom activities. Additionally, some teachers used the game to meet broader, school- or district-wide reading goals. For example, one teacher stated, "I needed to implement activities that would have a non-fiction connection to a story from our required grade level district mandated reading series." Similarly, another teacher stated that he used the game to support "nonfiction comprehension in reading class."

**Game as content container**
Some teachers viewed the instructional purpose of the game as a content container. In these instances, the game was used for its historical content: names, dates and themes. For example, one teacher stated that her main goal was to use the game to help students "identify important people events/facts about Wisconsin." After playing the game, the students researched Earth Day (which is one of the topic areas in the game), then wrote an informative paper and produced a poster to communicate a project the school could do related to Earth Day.  Another teacher said that his students would "use the *Jo Wilder and the Capitol Case* game to discover a little about Earth Day and its history in Wisconsin" and be able to "identify names of people that played a part of the first Earth Day." These goals are not surprising since they represent a prevailing view of what history is, namely, a collection of facts, dates and historical figures to be studied. Despite being designed to counter this approach toward teaching history, the game was framed by these teachers as a tool for exposing students to the historical content embedded in the game, rather than as a way to engage them in the process of doing historical inquiry.

**Games as a practice space**
Another instructional purpose for using *Jo Wilder* in the classroom was to provide a context for students to engage in disciplinary practices associated with history, such as collecting, interpreting, and corroborating evidence. One teacher captured this instructional purpose by stating that the game taught students to "think like a historian." In particular, many of the teachers commented on how the game required students to use historical thinking and reasoning in order to advance the story and solve the challenges. Along these lines, one teacher stated, "The students were better able to understand the process of historical inquiry after playing the game."

While playing the game afforded the opportunity for students to ask questions and make interpretations as they played, it is important to note that teachers did not simply rely on the game experience alone. One teacher emphasized how he called attention to what students did in the game after playing, so that they could reflect on the skills they were actually practicing. Another teacher developed a notebook that included some guiding questions to help students make sense of the artifacts they were looking at in the game and the skills they were using to interpret them.

**Games as a model**
Another instructional purpose we identified from the teachers' data was using the game as a model or case that could be used to reflect on what it means to engage in "doing history." For example, while one teacher had his students play *Jo Wilder* in order to be introduced to and practice historical inquiry, he also continued to reference the game as the students carried out historical inquiry on artifacts they brought in from home. In this way, he viewed the game as encapsulating a simplified version of "thinking like a historian." The teacher then structured the surrounding learning activities to mirror the kinds of activities *Jo Wilder* engaged in. He articulated that his unit "was designed so students would follow the same process *Jo Wilder* did while learning about historical items." He also explained that his class referred "to examples that Jo experienced in the game while we were conducting our own history investigations." This illustrates how the memory of the main character in the game and the in-game actions players took were intentionally linked to what they did in other parts of the unit. Another teacher used the game experience to have her students reflect on what it means to be a historian and make connections between historical inquiry and detective work. She also used the game as a model for discussing the use of evidence to make arguments and referenced the practices they did in the game when facilitating her students in their own inquiry work. In a similar vein, a couple of other teachers mentioned that they used the refrain "What would Jo do?" when working with their students as a way to draw attention to the practices she engaged in.

**Game as Launchpad**
A final instructional purpose utilized by the teachers was to use the game to launch or initiate a project-based instructional unit. As part of this instructional purpose, the game served to spark interest, introduce students to historical practices, and set the stage for them to do their own project. For example, one teacher had the students play *Jo Wilder* at the start of a unit as a way to introduce them to various aspects of the historical inquiry process. After playing the game, he led the class in a discussion to draw their attention to how Jo solved the cases in the game, such as by asking probing questions or taking notes. Next, students taught the game to second graders in their school as part of a class buddies program. The teacher specifically wanted his students to emphasize the skills that Jo was using and not just on completing the game. This all began with their initial launch into the game.

Another teacher used the game to introduce students to object-based inquiry. The game helped establish object-based inquiry as a mode and provided a structure for conducting historical inquiry. After playing the game, the teacher borrowed artifacts from the local historical society for students to interpret. In short, the game set the stage for the inquiry that the students engaged in throughout the project.

**Discussion**
Recognizing that teachers adapt materials to fit their goals, the needs of their students and other contextual factors challenges us to shift our attention away from prescribed uses of games towards a model that accounts for how they actually get used in classrooms. As Squire et al (2003) pointed out, curricular designs are not self-sufficient entities, but rather adapted, modified and carried out based on contextual factors. The instructional purposes described in our findings represent an emerging framework for how digital games get enacted in classrooms. While we do not claim that the five purposes we identified are comprehensive, we do see this framework as salient for supporting teachers implementation of game-based learning experiences. This paper does not claim to explain the different forms of enactment in direct relationship to the local contextual factors educators were working within. In future studies, we intend to investigate the extent to which the instructional purposes identified in this study are observed beyond this specific game and educator fellowship program. Are there additional forms of enactment? Are the forms found in this study simply local in nature and unique to this game and teacher group?

It is also important to note that we do not claim that any one instructional purpose we identified in this study is more effective than another. Instead, each purpose has its own benefits and drawbacks based on the respective goals, needs and values of the local classroom and teacher. For some, this might mean practicing reading with a variety of texts, including a digital game. For others, it might mean engaging in disciplinary practices with a digital game.

Identifying these instructional purposes, however, holds several potential benefits for the field. To begin, it acknowledges what educational researchers studying curriculum reforms have known for a long time; that is, teachers enact curriculum in their classrooms in different ways (Schneider et al, 2005; Bismack et al, 2014) and there is a difference between what is intended for curriculum and how it is actually facilitated in the classroom (Stein et al, 2007). Recognizing this phenomenon can help us design professional development that is more responsive to teachers' contexts and more in line with their actual problems of practice. In addition, it suggests that professional development might focus less on how to use a specific game and more on the adaptation process. Explicitly framing games in this light also reinforces the idea that games are resources that can be used to meet a wide range of goals, including ones not intended by the designers. Acting upon this perspective might expand teachers' (and pre-service teachers') chances of facilitating ambitious game-based teaching and learning experiences in their classroom.